\documentclass[aps,pra,twocolumn]{revtex4-1}

\usepackage{xifthen}
\usepackage{comment}
\usepackage[mathlines]{lineno}
\usepackage{wrapfig}
\usepackage{lipsum}
\usepackage{ulem}
\usepackage{graphicx}
\usepackage{amssymb}
\usepackage[colorlinks,allcolors=blue]{hyperref}
\usepackage[figure,figure*]{hypcap}
\usepackage{soul}
\usepackage{braket}
\usepackage{units}
\usepackage{xspace}
\usepackage{amsmath}

\mathchardef\mhyphen="2D

\newcommand{\nolinefrac}[2]{\genfrac{}{}{0pt}{}{#1}{#2}}

\newcommand{\pref}[2][]{\hyperref[#2]{\ref{#2}\ifthenelse{\isempty{#1}}{}{\sffamily(#1)}}}
\newcommand{\eqpref}[1]{\hyperref[#1]{(\ref{#1})}}

\newcommand{\Rb}{\textsuperscript{87}Rb\xspace}
\newcommand{\Na}{\textsuperscript{23}Na\xspace}
\newcommand{\Cs}{\textsuperscript{133}Cs\xspace}
\newcommand{\narb}{\texorpdfstring{\textsuperscript{23}Na\textsuperscript{87}Rb\xspace}{23Na87Rb}}
\newcommand{\rbcs}{\texorpdfstring{\textsuperscript{87}Rb\textsuperscript{133}Cs\xspace}{87Rb133Cs}}

\setstcolor{red}

\newcommand{\affTemple}{\affiliation{Department of Physics, Temple University, Philadelphia, PA 19122, USA}}

\newcommand{\affVT}{\affiliation{Department of Physics, Virginia Tech, Blacksburg, VA 24061, USA}}

\newcommand{\affUIUC}{\affiliation{Department of Physics and IQUIST, University of Illinois at Urbana-Champaign, Urbana, IL 61801-3080, USA}}

\begin{document}

\title{Nondestructive dispersive imaging of rotationally excited ultracold molecules}

\author{Qingze Guan}
\thanks{These authors contributed equally to this work.}
\affTemple

\author{Michael Highman}
\thanks{These authors contributed equally to this work.}
\affUIUC

\author{Eric J. Meier}
\affUIUC

\author{Garrett R. Williams}
\affUIUC

\author{Vito Scarola}
\email{scarola@vt.edu}
\affVT

\author{Brian DeMarco}
\email{bdemarco@illinois.edu}
\affUIUC

\author{Svetlana Kotochigova}
\email{svetlana.kotochigova@temple.edu}
\affTemple

\author{Bryce Gadway}
\email{bgadway@illinois.edu}
\affUIUC

\date{\today}

\begin{abstract}
A barrier to realizing the potential of molecules for quantum information science applications is a lack of high-fidelity, single-molecule imaging techniques. Here, we present and theoretically analyze a general scheme for dispersive imaging of electronic ground-state molecules. Our technique relies on the intrinsic anisotropy of excited molecular rotational states to generate optical birefringence, which can be detected through polarization rotation of an off-resonant probe laser beam. Using \narb and \rbcs as examples, we construct a formalism for choosing the molecular state to be imaged and the excited electronic states involved in off-resonant coupling. Our proposal establishes the relevant parameters for achieving degree-level polarization rotations for bulk molecular gases, thus enabling high-fidelity nondestructive imaging. We additionally outline requirements for the high-fidelity imaging of individually trapped molecules.
\end{abstract}

\maketitle

\section{Introduction}

Ultracold molecules are a promising platform for quantum information science (QIS) applications \cite{DeMille2002,Kuznetsova2012,Koch2019,DipolarKangKuen}. The abundance of long-lived rotational states in molecules is an advantage compared with using simpler quantum particles such as neutral atoms, for example. However, the lack of high-fidelity imaging techniques for general classes of molecules is a barrier to progress in this area.

For ultracold atoms and molecules, imaging plays a key role as the method for  state detection in a wide range of quantum control and information processing applications. For example, accurate readout of quantum processors based on trapped atomic ions requires high-fidelity imaging \cite{bruzewicz2019trapped}. Measurement via non-destructive, high-accuracy imaging is necessary to generate defect-free qubit registers in optical tweezer experiments \cite{Barredo1021,Endres1024}. Non-destructive imaging will also be critical to implementing alternative approaches such as measurement-based universal quantum computing \cite{Raussendorf2001b,Raussendorf2003} using atoms and molecules in the future. In this case, high-fidelity detection is also needed to realize fault tolerance ~\cite{Raussendorf2006,Raussendorf2007,Raussendorf2007b}.

In experiments with neutral atoms and atomic ions, high-fidelity imaging is achieved using closed transitions between ground and excited states. For these ``cycling" transitions, thousands of absorption and spontaneous emission events can occur before quantum amplitude leaks out of the manifold of imaging states. Certain classes of molecules with nearly diagonal Franck-Condon factors also possess quasi-closed cycling transitions \cite{DiRosa-cooling,Stuhl-MOT}. These transitions have paved the way for the direct laser cooling \cite{Shuman-Cooling}, trapping \cite{Barry-MOT}, high-fidelity fluorescence imaging of molecular samples \cite{Cheuk-Imaging}, and even individually trapped molecules \cite{doylecaf}. However, the internal state complexity of molecules due to rotational and vibrational degrees of freedom \cite{MoleculeReview-CarrDemilleKrems-Ye} generally preclude most molecules from possessing cycling transitions.

In particular, the bi-alkali molecules do not have closed or quasi-closed cycling transitions. Bi-alkali molecules can be readily prepared from pre-cooled atoms near \cite{Ni-HighPSD,Moses-LowEntropy-Molecules,HCN-lowentropy-molecules} or in the quantum degenerate regime \cite{DeMarco-Degenerate-Molecule} and have already demonstrated many-body physics \cite{Yan-DipDip-14,Hazzard-Gadway-PRL-14,Bloch-Mol-Exchange}. Direct imaging techniques for the bi-alkalis are presently based on absorption imaging using open, lossy optical transitions \cite{Wang-DirectImaging}. Alternatively, these molecules can be detected via the imaging of the constituent atoms following the coherent reversal of STImulated Raman Adiabatic Passage (STIRAP) \cite{Ni-HighPSD,Chotia-LongLived}. Both approaches are inherently destructive and lack the fidelity necessary for QIS applications.

In this paper, we present an alternative imaging technique that is applicable to a broad range of molecules, including the bi-alkalis. We propose to use the inherent anisotropic polarizability of rotationally excited molecules to allow nondestructive detection through birefringent phase shifts imparted on an off-resonant ``probe'' laser beam. We describe conditions under which degree-level polarization rotations of a probe beam can be achieved for bulk molecular gases, and we outline paths to extend this capability to the imaging of individual molecules. We construct a formalism for computing observable phase shifts and apply it to two example molecules, \narb and \rbcs. We identify specific states that optimize imaging resolution. Our calculations show that nondestructive imaging of birefringent phase shifts is within the reach of current technology.

The organization of the paper is as follows: In Section~\ref{sec_background} we discuss general aspects of dispersive imaging and propose a setup to measure phase shifts. In Sections~\ref{sec_select_imaging} and ~\ref{sec_select_target} we discuss the criteria for selecting the imaging and target states, respectively. Sections~\ref{sec_summary_bulk} and \ref{sec_summary_single} summarize our findings for application to imaging of bulk molecular gases and individually trapped molecules, respectively. Finally, in Section~\ref{sec_discussion}, we review the main results of this paper and identify a few specific areas in which the dispersive imaging of molecules may have future impacts.

\section{Background on Dispersive Imaging}
\label{sec_background}
 
Dispersive imaging is based upon interference of two or more off-resonant laser beams that have acquired a relative phase due to their different propagation through an atomic \cite{Bradley-Dispersive,AndrewsDispersive} or molecular medium. While the two beams can propagate along distinct paths~\cite{Hecker-Imaging,Smits:20} or involve distinct spatial regions of a single probe beam~\cite{AndrewsDispersive,Wigley-Dispersive}, approaches based upon co-propagating polarization~\cite{Sherson-QND,takahashisingleatom} or frequency~\cite{Hardman:16} components have the benefit that they are simple, robust, and inherently afford significant
common-mode noise rejection. For atoms, which typically possess cycling transitions, dispersive imaging has proven especially useful for niche applications in which one does not want to disturb density or temperature, so as to allow for continuous monitoring of a sample~\cite{Wilson-insitu,Nguyen422}. 

For molecules that lack true cycling transitions, however, dispersive imaging may provide the best means to achieve high-fidelity imaging. 
Therefore, the development of such a technique has the potential to find more widespread use for bi-alkali molecules and other species, while still allowing for nondestructive imaging. 
Polarization-based dispersive imaging thus promises to leverage one of the characteristic qualities of molecules -- their anisotropic tensor polarizability~\cite{Neyenhuis-AnisPol} -- for high-fidelity imaging and internal state detection.

\begin{figure}
\centering
\includegraphics{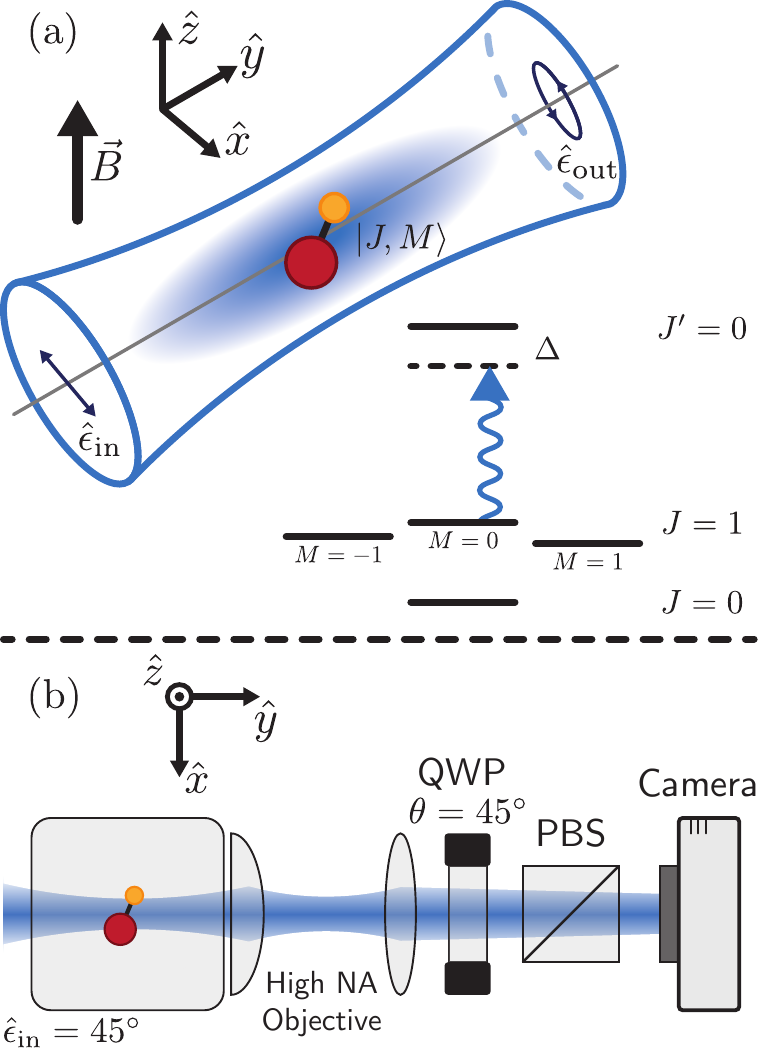}
\caption{Schematic of dispersive imaging setup for molecules. \textbf{(a)}~Molecules are illuminated by a probe beam propagating along the $\hat y$ direction, perpendicular to an external magnetic field $\vec B$ along the $\hat z$ direction. The $\hat z$ axis serves as the quantization axis. The probe laser polarization $\hat{\epsilon}_{\rm in}$ is linear and in the $x$-$z$ plane. The ellipticity of the output polarization $\hat{\epsilon}_{\rm out}$ depends both on $\hat{\epsilon}_{\rm in}$ and the rotational state $|J,M\rangle$ of the molecule(s). The probe beam in the perpendicular imaging case off-resonantly (with a frequency detuning of $\Delta$) couples an excited rotational state with primarily $M=0$ character to a $J'=0$ excited electronic state, as displayed in the level diagram. \textbf{(b)} One possible experimental setup for perpendicular imaging. Light with linear polarization at $45^\circ$ to the quantization axis acquires a differential phase shift through rotationally excited molecules. The light is collected by a high numerical aperture (NA) objective. The phase shift is translated to a polarization rotation via a quarter wave plate (QWP) with fast axis set at $45^\circ$ from vertical. A polarizing beam splitter (PBS) turns the rotation into a power difference that is detected by a camera.
}
\label{fig:cartoon1}
\end{figure}

Figure~\pref[a]{fig:cartoon1} schematically shows an example of a polarization-based setup for the dispersive imaging of molecules. A probe laser propagates through a molecular cloud along the $\hat y$ axis perpendicular to a uniform magnetic field $\vec{B}$ applied
along the $\hat z$ axis.  We call this the  ``perpendicular'' imaging scheme, in reference to the fact that the probe laser propagation and magnetic field direction are perpendicular.  For bi-alkali molecules that are first associated from atoms into molecules by means of a sweep across a Feshbach resonance, the magnetic field strength $B$ would 
typically be a few hundred Gauss, as determined by the Feshbach resonance. 
We consider the incident probe laser polarization $\hat{\epsilon}_{\rm in}$ as being linear and in the $x$-$z$ plane. If both the molecular rotational state $\ket{J,M}$ has an anisotropic dynamic polarizability tensor $\alpha_{\epsilon\epsilon'}(\omega)$,
for which the
indices $\epsilon$ and $\epsilon'$ are $x$, $y$, or $z$ in Cartesian coordinates, and $\hat{\epsilon}_{\text{in}}$ has a component both parallel and perpendicular to the quantization axis, then the output laser polarization $\hat{\epsilon}_{\text{out}}$ becomes elliptical.
Here, the rotational quantum number $J$ labels eigenstates of $\vec J$, 
the sum of the electronic and molecular-orbital angular momenta and $M$ is the projection along the quantization axis.
The phase difference $\phi(\omega)$ between the $z$- and 
$x$-components of the output laser beam
is given by 
\begin{eqnarray}
\phi(\omega)=\frac{2\pi\rho c L}{\lambda}\Delta\alpha(\omega), 
\label{eq:phi}
\end{eqnarray}
where $\lambda$ is the photon wavelength of the probe laser of angular frequency $\omega$, $\rho$ is the number density of the molecules, $L$ is the sample length, $c$ is the speed of light in vacuum, and $\Delta\alpha(\omega)=\alpha_{zz}(\omega) - \alpha_{xx}(\omega)$ is the differential polarizability. 
Equation~\ref{eq:phi} assumes a low differential index of refraction: $\Delta n = \rho c \Delta\alpha\ll1$.
It is also important to note that $\alpha$ is actually a modified polarizability \textit{volume}, where $\alpha=\alpha_{\text{SI}}/2\epsilon_0c$, $\epsilon_0$ is the vacuum permittivity, and $\alpha_{\text{SI}}$ is the polarizability in standard SI units. As will be conveniently used later in this paper, $\alpha/h$, where \textit{h} is Planck's constant, has units MHz/(W/cm\textsuperscript{2}), which is experimentally understood as the ac Stark shift at a given laser intensity. Equation~\ref{eq:phi} defines our key observable and therefore motivates evaluation of $\Delta\alpha(\omega)$. As we will see, large differential polarizability arises from anistropic states which we find in the $J=1$ manifold depicted in Fig.~\pref[a]{fig:cartoon1}.

Figure~\pref[b]{fig:cartoon1} shows a schematic of a proposed detection apparatus.  After passing through the molecular sample, the phase difference of the two polarization components of the probe beam is translated to a polarization rotation by the use of a quarter wave plate. The rotation is then translated to a probe power difference by, \textit{e.g.}, a polarizing beam splitter and a camera.

Alternative to this ``perpendicular'' probing scenario, the system can be probed using a linearly polarized laser beam propagating parallel to the magnetic field direction. This may be useful in certain contexts, \textit{e.g.}, as in the case of planar 2D samples resolved by a quantum gas microscope.  In this ``parallel'' imaging scheme, the phase shift is also given by Eq.~\ref{eq:phi} with $\Delta\alpha =\alpha_{++}(\omega) - \alpha_{--}(\omega)$ where the indices ``$+$'' and ``$-$'' indicate spherical tensor components of the dynamic polarizability tensor. We note that for this case in which the molecular sample displays circular birefringence, optical activity leads to a direct rotation of the probe beam's linear polarization.

The efficiency of the perpendicular and parallel imaging schemes are comparable, yet the distinction is critical as the orientation of the probe laser and quantization axis will determine the relevant states that display the largest anisotropy. A detailed derivation of Eq.~\ref{eq:phi} for the two probing schemes is given in Appendix~\ref{append1}. In what follows we will focus primarily on the perpendicular imaging scheme in the main text and reserve the discussion of the parallel imaging scheme to Appendices~\ref{append1} and \ref{appendparallel}.

A strong signal in an experimental setup will be induced by a large differential polarizability. For example, in a typical ultracold sample density of \unit[$\rho=10$\textsuperscript{12}]{cm\textsuperscript{-3}}, probe wavelength \unit[$\lambda=770$]{nm}, and a sample length
$L=\unit[30]{\mu\text{m}}$, the differential polarizability $\Delta\alpha/h$ must have a value of \unit[3.6]{MHz/(W/cm\textsuperscript{2})} to achieve a phase difference of $1^{\circ}$. As we will discuss, these magnitudes of $\Delta\alpha$ can be found near resonant electric dipole transitions from anisotropic ${J\ne0}$ rotational states of molecules. We note that even the $J=0$ rotational ground state may have an induced anisotropic polarizability, if the degeneracy of the $J'=1$ manifold's $M$ states is broken by an amount that is large compared to their natural linewidth. For imaging on narrow transitions, this can be accomplished by the application of an electric field for polar molecules, and potentially even by state-dependent ac Stark shifts.

We expect our scheme to be generally applicable to molecular states with large anisotropies in dynamic polarizability. Such states should appear for generic families of molecules. To make quantitative estimates we focus on states of two specific bi-alkali molecules. In the main text we focus on imaging $^{23}$Na$^{87}$Rb molecules occupying the $J=1$ rotational level of its ${v=0}$ vibrational level of the electronic ground state X$^1\Sigma^+$. We also discuss imaging for \rbcs in Appendix~\ref{append3} as another example of the applicability of our technique.

\section{Selection of Imaging States}
\label{sec_select_imaging}

The optimal imaging states have a large differential polarizability.  Since anistropy enhances differential polarizability, we search for states that are as anisotropic as possible. Specifically, we focus on the $J=1$ rotational manifold, and we look for the state with the highest occupation of the $M=0$ projection at the relevant magnetic field for the specific Feshbach resonance used in the molecule creation. For \narb molecules occupying the ${v=0},{J=1}$ ro-vibrational state of their electronic state X$^1\Sigma^+$ we are guided by recent work \cite{Wang2016} using a magnetic field strength of \unit[335.6]{G}.  We will show that the best imaging state at this field also happens to be lowest in energy.

As depicted in Fig.~\pref[a]{fig:cartoon1}, the ${J=1}$ rotational state of the ${v=0}$ ground-state molecule has  three projections ${M=-1,0,+1}$. The projection degeneracy is broken by hyperfine interactions between the two nuclear quadrupole moments and the rotation of the molecule as well as Zeeman interactions for the nuclear spins \cite{Aldegunde2009,APetrov2013,Li2017}. We denote the nuclear spins of \Na and \Rb by $\vec \imath_{\rm Na}$ and $\vec\imath_{\rm Rb}$, respectively. Both have quantum number, or value, of 3/2. Their projection quantum numbers along the magnetic field direction are $m_{\rm Na}$ and $m_{\rm Rb}$, respectively. For all interactions the sum $M_{\rm tot}=M+m_{\rm Na}+m_{\rm Rb}$ is a conserved quantity. We use the nuclear quadrupole moments and nuclear $g$ factors from Refs.~\cite{Wang2016, Jesus2017}. Coupling to rotational states $J\ne 1$ is negligible as the rotational constant \cite{Guo2018} is orders of magnitude larger than the energy scales of the hyperfine and Zeeman interactions.

\begin{figure}
\includegraphics[scale=.6]{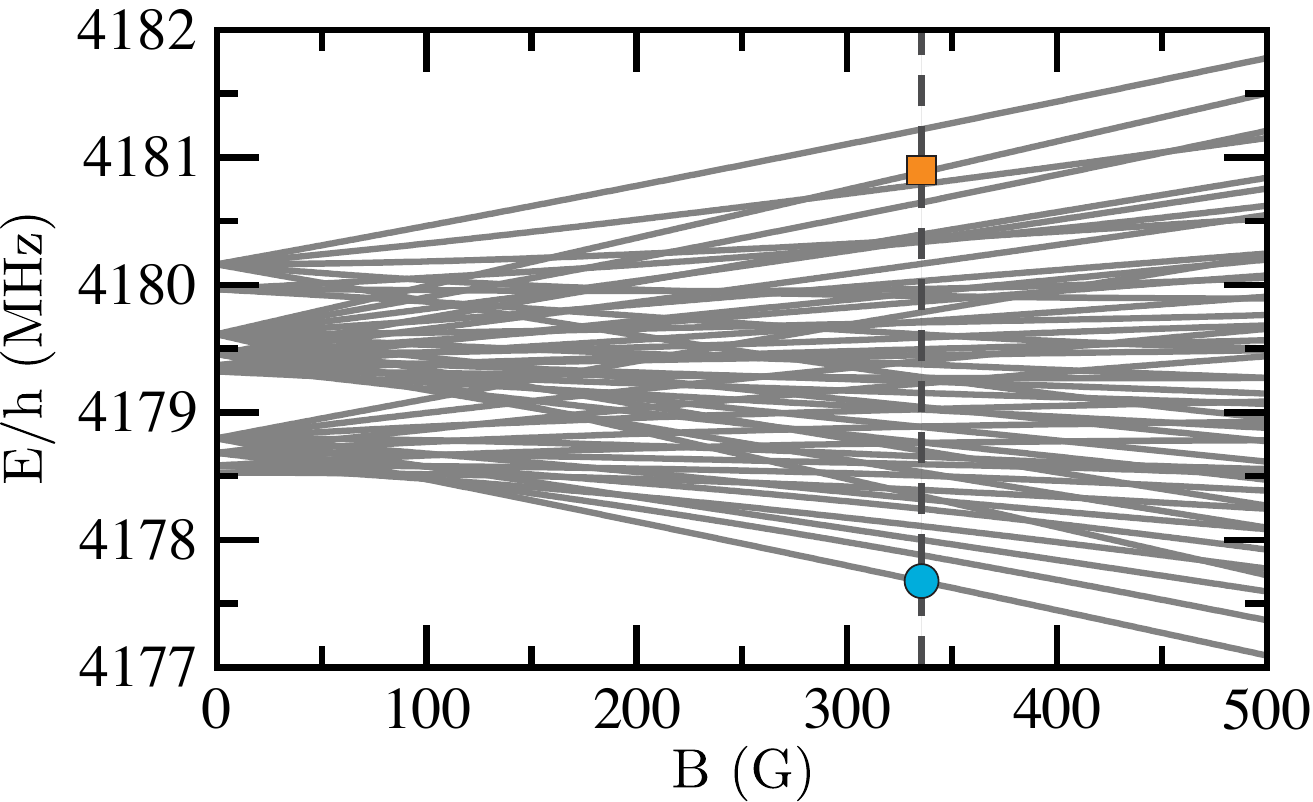}
\caption{Eigenenergies of the hyperfine states of the $v=0,\,J=1$ ro-vibrational level of the X$^1\Sigma^+$ electronic ground state of \narb as a function of magnetic field strength $B$. The dashed vertical line indicates the magnetic field strength \unit[$B=335.6$]{G}. The cyan dot and orange square mark states used for the perpendicular and parallel (Appendix~\ref{appendparallel}) imaging schemes, respectively. The zero of energy of this plot relates to the zero-field ($B=0$) energy of the ${v=0, J=0}$ level with no electron-nuclear quadrupole interaction.
}
\label{fig:zeeman}
\end{figure} 
  
There are 48 hyperfine-Zeeman eigenstates of the $v=0,\,J=1$ level of ground state \narb. In zero magnetic field the total angular momentum $\vec F_{\rm tot}=\vec J + \vec\imath_{\rm Na}+ \vec\imath_{\rm Rb}$ is conserved and states can also be labeled by $F_{\rm tot}$ as well as $M_{\rm tot}$. For magnetic field strengths larger than about \unit[100]{G} the nuclear Zeeman interaction is stronger than the hyperfine interactions and states with the same $M_{\rm tot}$ avoid each other. There, the energetically lowest ${J=1}$ state has ${M_{\rm tot}=+3}$. For $B$ fields not exceeding \unit[500]{G} the 48 levels span an energy range of no more than \unit[$h\times5$]{MHz}. Fig.~\ref{fig:zeeman} plots the relevant eigenenergies.  

For perpendicular dispersive imaging we investigate the lowest energy state depicted in Fig.~\ref{fig:zeeman} (cyan dot). We check that the hyperfine state has a relatively high component of the projection quantum number $M=0$ and relatively small contribution of other $M$ projections. Our calculations show that the energetically lowest ${J=1}$ level has the largest ${M=0}$ contribution and to very good approximation is described by the superposition
\begin{eqnarray}
|\varphi_{\rm perp}\rangle &=&  c_0
\left|\text{X}^1\Sigma^{+}; \nolinefrac{v=0,J=1,M=0}{m_{\text{Na}}=3/2, m_{\text{Rb}}=3/2} \right\rangle
\label{eq:imag_state_perp}  \\
  && ~+ \,c_1\left|\text{X}^1\Sigma^{+}; \nolinefrac{v=0, J=1,M=1}{m_{\text{Na}}=3/2, m_{\text{Rb}}=1/2}\right\rangle
 \nonumber
\end{eqnarray}
with coefficient $c_0 = 0.892$ and $c_1 = 0.452$. We will therefore proceed with this state as the imaging state.  

\section{Selection of target excited states}
\label{sec_select_target}

\subsection{Selection Criteria and Relevant Quantities}
\label{sec_criterea}

We aim to select target excited states that satisfy three criteria.  First and foremost the dynamic polarizabilty should display a large anisotropy near the resonance transition to the target excited state.  This will ensure detectability via large phase differences in Eq.~\ref{eq:phi} for our imaging state,  $\ket{\varphi_{\text{perp}}}$. Secondly, the target state, $|\psi_{\rm t}\rangle$, should have a small natural linewidth in order to minimize heating, particle loss, and dephasing. We also impose a third criteria as a matter of practical experimental concern. We additionally search for a target state where the transition has as large a transition width as possible, thus allowing for easier laser stabilization as well as more robust operation. This section defines the quantities we need: the natural linewidth, dynamical polarizability and photon scattering rate, to search for useful target excited states based on the above criteria.

We first consider the natural linewidth, $\gamma_{\rm n}$, of the target state \cite{scully1997,Romain2017}:
\begin{equation}
\label{eq:decay}
\gamma_{\rm n}=
\frac{4}{3} \frac{1}{4\pi\epsilon_0\hbar c^3}\sum_{g\vec{\epsilon}} \omega_{{\rm t} \mhyphen g}^3
|\langle\psi_{\text{t}}|d_{t\leftarrow g}(R) \hat{R}\cdot \vec{\epsilon}|\varphi_g \rangle|^2\,.
\end{equation}
Here the sum of $\vec\epsilon$ is over the  polarization direction of the spontaneously emitted photon and  the summation $g$ for hetero-nuclear alkali-metal dimers is over all eigenstates $|\varphi_g\rangle$, both bound and scattering states, with energy $E_g$ of the X$^1\Sigma^{+}$ and a$^3\Sigma^{+}$ potentials. Both potentials dissociate to atoms in the electronic ground state. The transition energy reads $\hbar\omega_{{\rm t}\mhyphen g}=E_{\rm t}-E_{g}$, where $E_{\rm t}$ is the energy of the  target state. The quantity $d_{t\leftarrow g}(R)$ is the $R$-dependent transition electric dipole moment operator, where $R$ is the interatomic separation. The interatomic axis has orientation $\hat R$.

We find it convenient to define the orientation-dependent ``transition widths'' $\Gamma_{\epsilon\epsilon}$ for transitions between the imaging state $|\varphi_{\text{perp}}\rangle$ and target state $|\psi_{\text{t}}\rangle$ using probe polarization $\vec \epsilon$ as:
\begin{eqnarray}
\Gamma_{\epsilon\epsilon}&=&
\frac{4}{3} \frac{1}{4\pi\epsilon_0\hbar c^3}
\omega_{\rm t\mhyphen im}^3 
\label{eq:resonance_width}
|\langle\psi_{\text{t}}|d_{t\leftarrow g}(R) \hat{R}\cdot \vec{\epsilon}\,|\varphi_{\text{perp}}\rangle|^2.
\end{eqnarray}
Here, $\hbar\omega_{\rm t\mhyphen im}=E_{\rm t}-E_{\rm im}$ and $E_{\rm im}$ is the eigenenergy of the imaging state. Since the imaging state $|\varphi_{\text{perp}}\rangle$ is a bound state of the X$^1\Sigma^{+}$ potential, it is thus included in the sum over states in Eq.~\ref{eq:decay}. All else held equal, target states with as large a value of $\Gamma_{\epsilon\epsilon}$ as possible may be practically desirable, as transitions to these states will be less sensitive to laser noise and technical variations of the state energies.

To highlight the anisotropy in $\Gamma_{\epsilon\epsilon}$ we define the differential transition width:
\begin{equation}
\Gamma=\Gamma_{zz}-\Gamma_{xx}
\label{eq_diff_linewidth}
\end{equation}
for $\Delta\alpha(\omega)$. We argue (See Appendix~\ref{append_derive_gamma}) that for our particular choice of target excited states, $\Gamma$ fully captures the anisotropy. For the $n'$-th~ro-vibrational target states we use here, $|\psi_{{\rm t},n'}\rangle$, we find (See Appendix~\ref{append_derive_gamma}):  
\begin{eqnarray}
\label{eq:transition_width_2}
\Gamma = \frac{4}{3} \frac{1}{4\pi\epsilon_0 \hbar c^3}
\left(\frac{|c_0|^2}{3} - \frac{|c_1|^2}{6}\right) \omega_{\rm t\mhyphen im}^3
 |\mu_{n'}|^2\,,
\end{eqnarray}
where the vibrational matrix elements $\mu_{n'}$ depend on the target state and are defined explicitly in Appendix~\ref{append_derive_gamma}. Here $n'=0,1,2,...$ is used to label the eigenstates by order of their eigenenergies. 

We now turn to the dynamic polarizability. The dynamic polarizability tensor components $\alpha_{\epsilon\epsilon}(\omega)$ of the imaging state $|\varphi_{\text{perp}}\rangle$ at probe frequency $\omega$ are determined by a sum over ro-vibrational levels and scattering states of all electronic states. For frequencies close to the target state resonance, such that $\omega\approx \omega_{\rm t\mhyphen im}$ but $|\omega - \omega_{\rm t\mhyphen im}|\gg \gamma_n$, the polarizability can be described as
\begin{eqnarray}
    \alpha_{\epsilon\epsilon}(\omega) &=& -\frac{3\pi}{2}\frac{ c^2}{\omega_{\rm t\mhyphen im}^3}
    \frac{\Gamma_{\epsilon\epsilon}}{\Delta}+\alpha_{\epsilon\epsilon}^{(0)},
    \label{eq:polarres}
\end{eqnarray}
where $\Delta=\omega-\omega_{\rm t\mhyphen im}$ is the probe laser detuning. The background polarizability $\alpha_{\epsilon\epsilon}^{(0)}$ contains the contributions from all other far-detuned molecular states and for our purposes can be taken as independent of $\omega$. We note that a similar background contribution to the polarizability anisotropy, $\Delta\alpha(\omega)$, can also be defined. This background anisotropy is several orders of magnitude smaller than the \unit[]{MHz/(W/cm\textsuperscript{2})}-level contributions we consider near resonance, and in practice can be safely neglected. We seek to find states where the difference of two components of this polarizability tensor, $|\alpha_{zz}-\alpha_{xx}|$, is maximized for a fixed detuning. Such an anisotropy can be achieved by looking for transitions with significant angular dependence of $\Gamma_{\epsilon\epsilon}$.

Finally, we will also compute the photon scattering rate to estimate heating and loss of coherence near a resonance. For $|\Delta|\gg\gamma_{\rm n}$ it is given by:
\begin{equation}
   \gamma_{\rm sc} =  \sum_{\epsilon\epsilon} \left(\frac{3\pi}{4} \frac{ c^2}{\omega_{\rm t\mhyphen im}^3}  \frac{\Gamma_{\epsilon\epsilon}\gamma_{\rm n} }{\Delta^2} 
     + \beta_{\rm sc}^{(0)} \right)\times I \ ,
     \label{eq:scatrate}
\end{equation}
where $I$ is the probe laser intensity and $\beta_{\rm sc}^{(0)}$ the background imaginary polarizability. Minimal values of $\gamma_{\rm sc}$ are ideal to avoid heating and scattering loss into dark states. 

\subsection{Target Excited States for \narb}

In this section we will show that starting from the imaging state (a hyperfine state of the ${v=0,J=1}$ ro-vibrational level of the X$^1\Sigma^{+}$ state) we can use optical wavelengths to access mixed $J'=0$ ro-vibrational states of the coupled A$^1\Sigma^{+}$-b$^3\Pi_0$ complex. We will show by direct calculation that these states satisfy the criteria discussed in Sec.~\ref{sec_criterea}.

\begin{figure}
\includegraphics[scale=0.38]{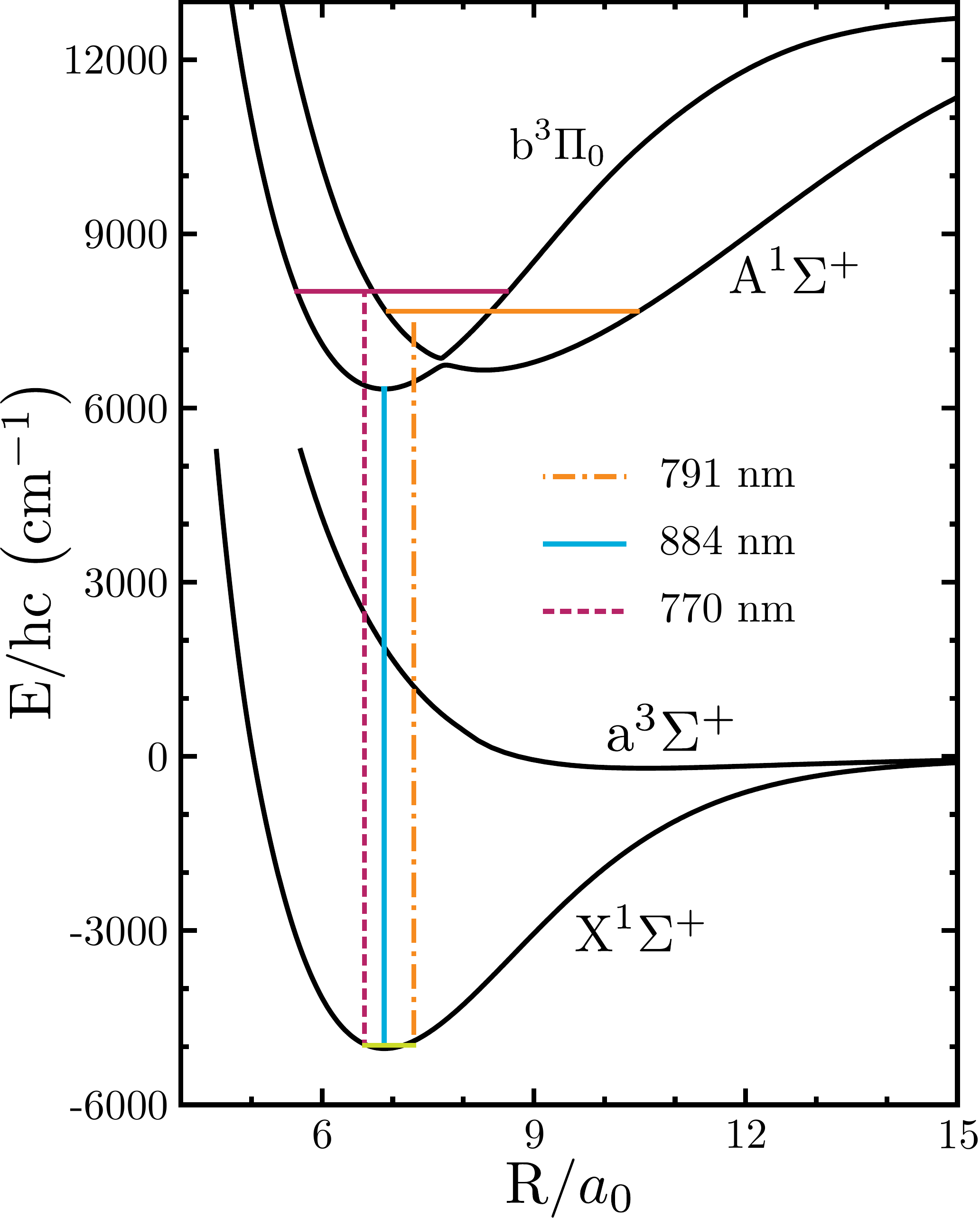}
\caption{Ground and relevant excited adiabatic relativistic potentials of the \narb molecule as functions of atom-atom separation $R$. The two energetically lowest adiabatic potentials are identified by non-relativistic labels X$^1\Sigma^+$ and a$^3\Sigma^+$ respectively. The zero of energy of the graph is set at their dissociated limit. The two remaining excited adiabatic potentials have a narrow avoided crossing at $R_{\rm c}\approx7.5a_0$. For $R>R_{\rm c}$ the electronic wavefunctions of the third and fourth adiabat are well described by the non-relativistic A$^1\Sigma^{+}$ and b$^3\Pi_0$ symmetry, respectively. For $R<R_{\rm c}$ this assignment is inverted. The three vertical lines indicate transitions from the $v=0,J=1$ imaging state in the X$^1\Sigma^+$ state to three mixed $J'=0$ ro-vibrational states
of the coupled A$^1\Sigma^{+}$-b$^3\Pi_0$ complex. The transition wavelengths are \unit[770]{nm}, \unit[791]{nm}, and \unit[884]{nm} for the magenta, orange, and cyan lines, respectively. The magnetic field is \unit[$B=335.6$]{G}.
}
\label{fig:pots}
\end{figure} 

We now present results for our exhaustive search for useful target states. Fig.~\ref{fig:pots} effectively summarizes the findings of this section by plotting the transition to the relevant target states against the molecular potential. The details in obtaining these target states can be found in Appendix~\ref{appendix_new}. Our search led to a focus on three states highlighted in Fig.~\ref{fig:pots}. To select target states that are convenient for imaging we have computed the natural linewidth and differential transition width for all eigenstates of the $J'=0$ A$^1\Sigma^{+}$-b$^3\Pi_0$ system. The lowest energy excited level is to the $n'=0$ eigenstate. It has a \unit[99.75]{\%} admixture $a_{b,n'=0}$ in the $|\text{b}^3\Pi_0\rangle$ state. A transition from an imaging state to this target state has a wavelength of \unit[884]{nm}. The transition from our imaging state to the $n'=29$ target eigenstate has a wavelength of \unit[791]{nm}. This target state has a \unit[96.85]{\%} admixture in the $|\text{A}^1\Sigma^{+}\rangle$ state. Finally, the \unit[770]{nm} transition is to the $n'=39$ eigenstate. This target state has a \unit[94.65]{\%} admixture in the $|{\rm b}^3\Pi_0\rangle$ state and was used in Ref.~\cite{Wang2016} as the intermediate state in the STIRAP process to form \narb molecules in their absolute ground state.

\begin{figure}
\includegraphics[scale=0.6]{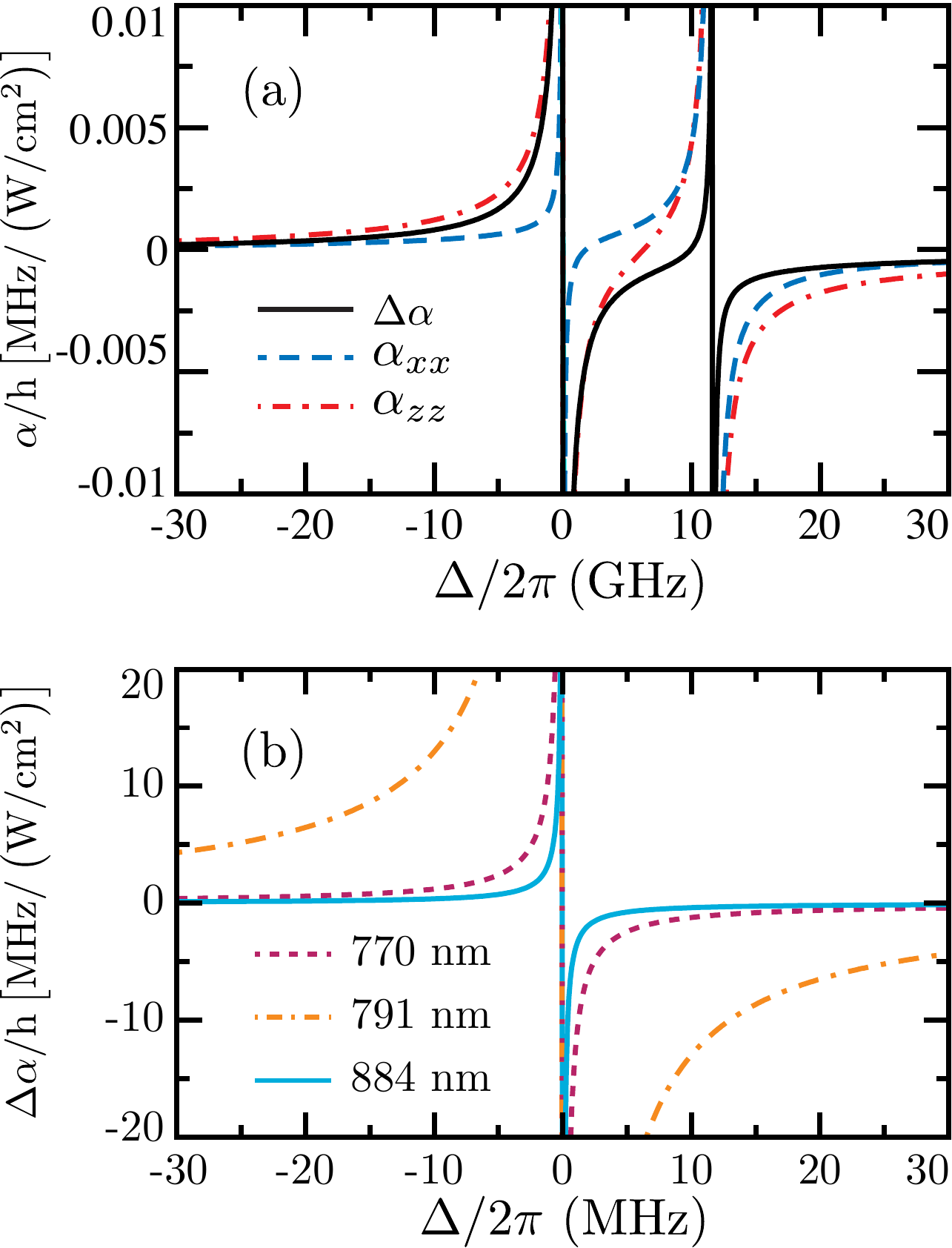}
\caption{\textbf{(a)} Values of $\alpha_{xx},\,\alpha_{zz},\,\text{and}\,\Delta\alpha$ for a large range of detunings of the \unit[770]{nm} transition from the ground $v=0,J=1$ X$^1\Sigma^+$ imaging state $|\varphi_{\rm perp}\rangle$. The second resonance at roughly \unit[12]{GHz} corresponds to transitions to the $J'=2$ states. \textbf{(b)} Differential dynamic polarizabilities $\Delta\alpha$ as a function of frequency detuning $\Delta$ to the target states for the \unit[770]{nm}, \unit[791]{nm}, and \unit[884]{nm} transitions of \narb identified in Fig.~\ref{fig:pots}. The magnetic field in both graphs is \unit[335.6]{G}.
}
\label{fig:deltaalpha}
\end{figure} 

Figure~\pref[a]{fig:deltaalpha} shows the components of the differential dynamic polarizability of $|\varphi_{\rm perp}\rangle$ for the \unit[770]{nm} transition as a function of detuning $\Delta$. Visible are the poles of both a $J'=0$ and $J'=2$ transition. Since the state $|\varphi_{\rm perp}\rangle$ has an \unit[80]{\%} and \unit[20]{\%} population in the $M=0$ and $M=1$ components, respectively, the transition width $\Gamma_{xx}$ of $\alpha_{xx}(\omega)$ is much smaller than the corresponding width of $\alpha_{zz}(\omega)$. Due to these unbalanced populations, the differential transition width $\Gamma$ is positive for negative detuning, hence the differential dynamic polarizability $\Delta\alpha$. On the vertical scale of the figure the background contribution $\Delta\alpha^{(0)}$ to $\Delta\alpha(\omega)$ is negligible. Narrowing in on the $J'=0$ transition, we summarize in Fig.~\pref[b]{fig:deltaalpha} the results for $\Delta\alpha$ for the \unit[770]{nm}, \unit[791]{nm}, and \unit[884]{nm} transitions. Here we see the resonant transition near \unit[791]{nm} to target state $n^{\prime}=29$ with its large A$^1\Sigma^{+}$ admixture has the largest differential transition width $\Gamma$ by far. This is a consequence of the large transition dipole moment between the X$^1\Sigma^{+}$ and A$^1\Sigma^{+}$ states. Naively, this suggests that this transition is the best of the three candidate transitions for perpendicular imaging. We, however, must also account for spontaneous emission and, in particular, whether the photon scattering rate $\gamma_{\text{sc}}$ is minimized.

\begin{figure}
\includegraphics[trim = 20 45 0 80, clip,width=1\textwidth]{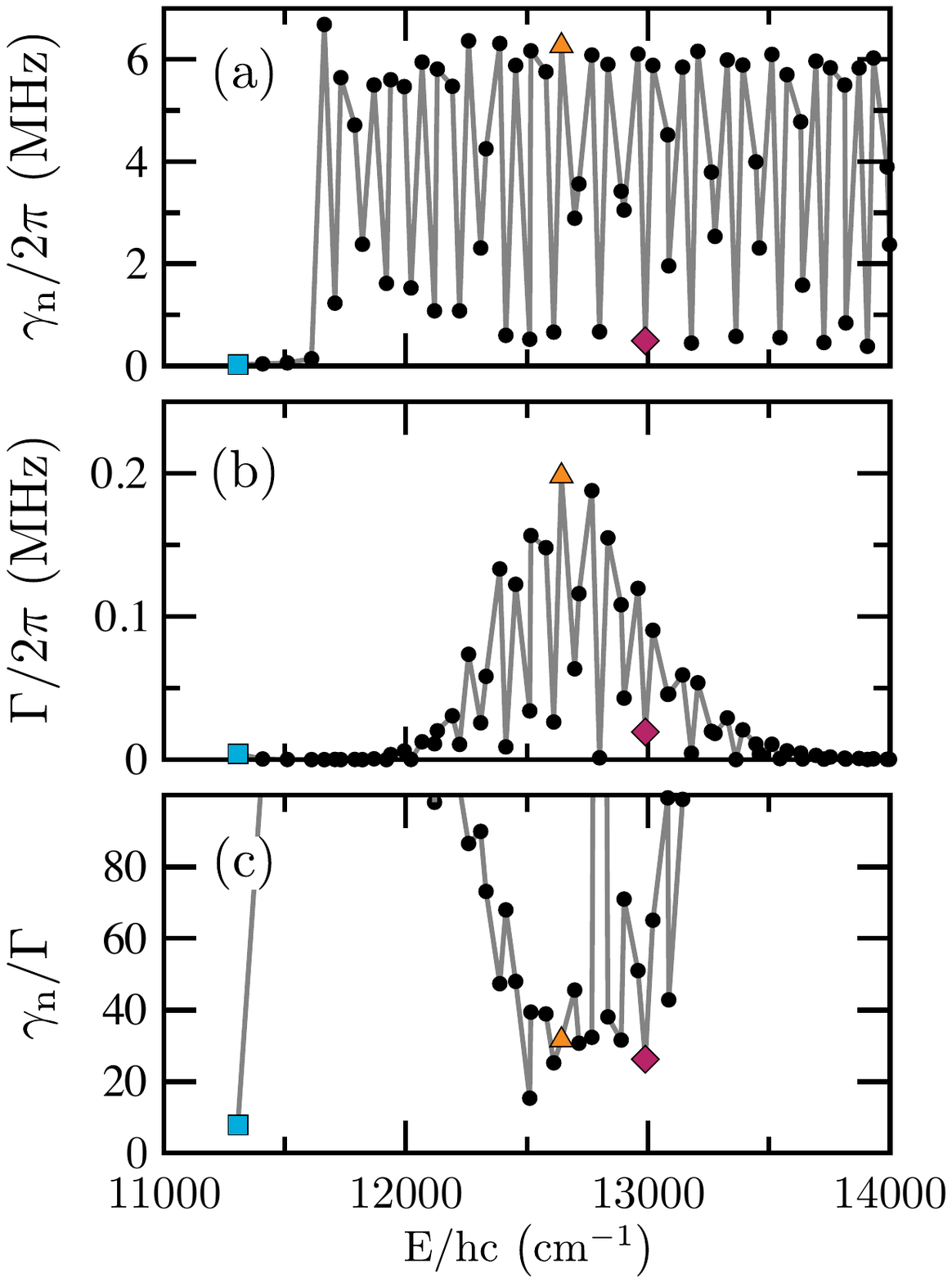}
\caption{The natural linewidth $\gamma_{\rm n}$ \textbf{(a)}~of eigenstates of the $J'=0$ A$^1\Sigma^{+}$-b$^3\Pi_0$ complex of \narb and their differential transition width $\Gamma$ \textbf{(b)}~from the ground $v=0,J=1$ X$^1\Sigma^+$ imaging state $|\varphi_{\rm perp}\rangle$ for perpendicular imaging as functions of transition energy $E$. The transition energies are relative to the energy of $|\varphi_{\rm perp}\rangle$. \textbf{(c)}~The ratio $\gamma_{\rm n}/\Gamma$ as a function of transition energy. The cyan square, orange triangle, and magenta diamond correspond to the transitions shown in Fig.~\ref{fig:pots} featuring transition wavelengths of \unit[884]{nm}, \unit[791]{nm}, and \unit[770]{nm}, respectively. The values of $\gamma_{\rm n}/\Gamma$ for the cyan square, the orange triangle, and the magenta diamond are 7.8, 32, and 26, respectively. The magnetic field in all graphs is \unit[335.6]{G}.
}
\label{fig:GammaNaRb}
\end{figure}

To look for the transition with the best balance between large transition width and small photon scattering rate, we have additionally determined $\Delta\alpha(\omega)$ for $|\varphi_{\rm perp}\rangle$ as a function of $\omega$ near transitions to many of the $J'=0$ eigenstates of the A$^1\Sigma^{+}$-b$^3\Pi_0$ complex.  We have also computed the natural linewidths and differential transition widths, $\gamma_{\rm n}$ and $\Gamma$, of these target states. Fig.~\ref{fig:GammaNaRb} shows widths $\gamma_{\rm n}$ and $\Gamma$ as well as the ratio $\gamma_{\rm n}/\Gamma$ for the first 66 $J'=0$ eigenstates of the A$^1\Sigma^{+}$-b$^3\Pi_0$ complex. The colored markers in each panel correspond to the three transitions shown in Figs.~\ref{fig:pots} and \pref[b]{fig:deltaalpha}. The left-most four points with the smallest transition energy correspond to transitions to the bound states at the bottom of the $\text{b}^3\Pi_0$ potential.

Figure~\pref[a]{fig:GammaNaRb} shows that the natural linewidths $\gamma_{\rm n}$ group roughly into three bands: those with values smaller than \unit[2$\pi\times$2]{MHz}, those with values larger than \unit[2$\pi\times$5]{MHz}, and those in between. The first corresponds to transitions to target states with a dominant admixture of the $|{\rm b}^3\Pi_0\rangle$ state and thus would have been forbidden without spin-orbit coupling between the A$^1\Sigma^{+}$ and b$^3\Pi_0$ states. The second group corresponds to transitions to target states with a dominant admixture in the A$^1\Sigma^{+}$ state leading to the largest $\gamma_{\rm n}$. Finally, the scattered points between these two bands correspond to target states with almost equal admixture of b$^3\Pi_0$ and A$^1\Sigma^{+}$ components. The natural linewidths for the \unit[884]{nm}, \unit[791]{nm}, and \unit[770]{nm} transitions are calculated to be \unit[2$\pi\times$0.027]{MHz}, \unit[2$\pi\times$6.3]{MHz}, and \unit[2$\pi\times$0.50]{MHz}, respectively. 

Figure~\pref[b]{fig:GammaNaRb} shows the differential transition widths $\Gamma$. Their values are positive, oscillate with transition energy, and have a Gaussian envelope. For a transition with larger $\Gamma$, we have a larger absolute range of detunings in which the differential polarizability can reach the desired magnitude. In fact, $\Gamma$ is largest when the target state has a large A$^1\Sigma^+$ admixture and the vibrational matrix element $\mu_{n'}$ is large. The latter occurs when the inner turning point of the vibrational motion on the A$^1\Sigma^+$ potential coincides with the equilibrium separation of the X$^1\Sigma^+$ potential. The \unit[791]{nm} transition to the $n'=29$ A$^1\Sigma^{+}$-b$^3\Pi_0$ eigenstate, already discussed in the context of Figs.~\ref{fig:pots} and \pref[b]{fig:deltaalpha}, has the largest $\Gamma$. Finally, we observe that the differential transition widths for the \unit[884]{nm}, \unit[791]{nm}, and \unit[770]{nm}, transitions are \unit[2$\pi\times$4.0]{kHz}, \unit[2$\pi\times$190]{kHz}, and \unit[2$\pi\times$19]{kHz}, respectively.

Figure~\pref[c]{fig:GammaNaRb} shows the ratio $\gamma_{\rm n}/\Gamma$ as a function of transition energy from the $|\varphi_{\rm perp}\rangle$ state. This quantity gives insight to the ``verticality'' of the transition, wherein smaller ratios correspond to the fewest decay paths available to the targeted excited state. For example, for a target state that can only spontaneously decay to the ground state (same electronic, vibrational, and hyperfine levels as the imaging state),
%imaging state $|\varphi_{\rm perp}\rangle$
we find the lower bound for this ratio is
\begin{equation}
\frac{\gamma_{\rm n}}{\Gamma}= \frac{3}{|c_0|^2-|c_1|^2/2 }.
\label{eq:ratio_limit}
\end{equation}
For the state $|\varphi_{\rm perp}\rangle$, this limiting ratio is 4.3. A lower bound ratio of 3 is found in the ideal case of a pure $|J = 1,0 \rangle$ state for perpendicular imaging, limited by the $J'=0$ target state's ability to decay to any of the three $M$ states of the $J=1$ manifold. Transitions that realize this lower bound are known as vertical transitions. For the case of a $J=0$ imaging state with anisotropic polarizability induced by, \textit{e.g.}, an applied electric field,
%to break the degeneracy of the $J' = 1$ sublevels,
the lower bound ratio for vertical transitions is also equal to $3$, due to the dipole-allowed decay paths to $J = 0$ and 2 states.

The ratio $\gamma_{\text{n}}/\Gamma$ is larger than 10 for all transitions to $J'=0$ A$^1\Sigma^{+}$-b$^3\Pi_0$ eigenstates except for the $n'=0$ eigenstate, where its value is 7.8. Thus this \unit[884]{nm} transition is closest to vertical. For all the transitions, the excited state does not only spontaneously decay to the imaging state but also to other ro-vibrational states of the X$^1\Sigma^{+}$ potential as well as those of the a$^3\Sigma^+$ potential. The most typical value for $\gamma_{\rm n}/\Gamma$ is between 20 and 40. Lastly, we calculate $\gamma_{\text{n}}/\Gamma$ values of 32 and 26 for the \unit[791]{nm} and \unit[770]{nm} transitions, respectively. We note that these values are all reduced by a factor of 1.44 for the case of parallel imaging.

The next section summarizes how these results for $\gamma_{\text{n}}/\Gamma$ and $\gamma_{sc}$ relate to an interplay and trade-off with respect to maintaining low inelastic scattering rates and allowing for robust operation.

\section{Summary of perpendicular imaging conditions for bulk gases}
\label{sec_summary_bulk}

We have shown that polar molecules prepared in rotationally excited states can act as an anisotropic medium, resulting in birefringent phase shifts on an off-resonant probe laser field. Furthermore, our calculations show that these phase shifts are large enough to be detectable. For the three transitions identified in \narb in the previous section, we summarize in Table~\ref{tab:1} the detuning, in units of the respective transition linewidth, necessary to achieve a birefringent phase shift $\varphi_{\rm{bulk}}=1^{\circ}$ and the resulting inelastic loss rate $\gamma_{\rm{sc}}$. Because the ratios of the natural linewidth to the transition linewidth, $\gamma_{\text{n}}/\Gamma$, differ for the various excited states considered, we see a range of detunings that are necessary to attain the $1^{\circ}$ polarization rotation. Here, we have considered typical density values, \unit[$\rho=10^{12}$]{cm\textsuperscript{-3}}, for molecular gases formed from pre-cooled atoms~\cite{Ni-HighPSD,DeMarco-Degenerate-Molecule}, and a sample length $L$ equal to \unit[30]{$\mu$m} (long, but readily achievable for single-beam trapping). For the inelastic loss rates presented in Table~\ref{tab:1}, we have considered a probe beam intensity of \unit[0.02]{mW/cm\textsuperscript{2}} (relating to the peak probe intensity for a beam with \unit[50]{$\mu$W} of power and a \unit[1]{inch} diameter).

\begin{table}
\begin{tabular}{|c|c|c|c|c|}
    \hline
    Wavelength & $1^{\circ}\,\Delta\,(\Gamma)$ & \unit[$1^{\circ}\,\Delta/2\pi$]{(MHz)} & $\gamma_{\text{n}}/\Gamma$ & \unit[$\gamma_{\text{sc}}/2\pi$]{(Hz)} \\
    \hline
    \unit[884]{nm} & 159 & 0.64 & 7.8 & 4.08   \\
    \hline
    \unit[791]{nm} & 134 & 25.43 & 32 & 16.91  \\
    \hline
    \unit[770]{nm} & 124 & 2.35 & 26 & 14.85\\
    \hline
\end{tabular}
\caption{Summary of relevant quantities for the three chosen transitions of the A$^1\Sigma^{+}$-b$^3\Pi_0$ system of \narb. The first column gives the transition wavelength. The second and third columns give the detuning, in units of transition linewidths and MHz, respectively, necessary to attain a $1^{\circ}$ polarization rotation on a molecular sample of density \unit[10\textsuperscript{12}]{cm\textsuperscript{-3}} and a sample length of \unit[30]{$\mu$m}. The fourth column shows the ratio of the natural linewidth to the differential transition width. The last column is the inelastic scattering rate $\gamma_{\rm sc}$ when achieving $1^{\circ}$ rotations for a probe intensity of \unit[0.02]{mW/cm\textsuperscript{2}}. All quantities relate to the case of ``perpendicular'' imaging. In the case of ``parallel'' imaging, $\gamma_{n}/\Gamma$ is reduced by a factor of 1.44 and the inelastic scattering rate $\gamma_{sc}$ is reduced by a factor of 1.87 for an equivalent rotation angle $\phi$.
}
\label{tab:1}
\end{table}

In the previous section, advantages of choosing the target states corresponding to the \unit[791]{nm} and \unit[884]{nm} transitions were briefly discussed. The \unit[791]{nm} transition is the strongest yet it maintains a reasonably small $\gamma_{\text{n}}/\Gamma$ ratio. These work to keep the inelastic scattering rate low while reducing the amount of laser stability needed to maintain a particular value of detuning (in units of $\Gamma$). The \unit[884]{nm} transition, albeit much weaker, has the lowest $\gamma_{\text{n}}/\Gamma$ ratio at 7.8 and is therefore subject to the smallest amount of imaging induced heating~\footnote{For each of the identified transitions for \narb, if one considers the parallel imaging scheme as compared to perpendicular imaging, the ratio of $\gamma_{n}/\Gamma$ is lower by a factor of 1.44 and the inelastic scattering rate $\gamma_{sc}$ at an equivalent rotation angle $\phi$ is lower by a factor of 1.87}. Because of the narrow differential transition width of \unit[884]{nm} transition, its requirements for laser stabilization and its sensitivity to noise and drifts of the state energies will be more pronounced. However,
%that none of the identified transitions are ultra-narrow~\cite{Koto-Bloch-Narrow}
because this dispersive imaging scheme can be operated dozens or hundreds of differential transition widths away from resonance, it is in general rather insensitive to such frequency variations.

The \unit[770]{nm} transition sits at a compromise, in both transition strength and $\gamma_{\text{n}}/\Gamma$ ratio, between the \unit[791 and 884]{nm} transitions. The primary benefit is that experiments with ground state \narb will necessarily have the laser stabilization infrastructure for this wavelength in place, as it is used in the production of ground state molecules by STIRAP. As the STIRAP ``dump'' (Stokes) laser is typically fixed to the $J = 0 \rightarrow J' = 1$ transition frequency by locking to a cavity by the Pound-Drever-Hall (PDH) method, a stable imaging beam detuned by $\Delta$ from the $J = 1 \rightarrow J' = 0$ transition may easily be engineered without the need for an additional stabilized laser. This could be accommodated by using acousto-optic modulators to introduce GHz-level frequency shifts ($2B \pm \Delta \sim h\times 4$~{GHz} for \narb), or by dynamically changing the frequency offset used for PDH sideband locking (in the case that a broadband fiber electro-optic modulator can be utilized) prior to imaging. Given the availability of suitable imaging light in \narb experiments~\cite{Wang2016}, the realization of nondestructive dispersive imaging of \narb molecular gases should be imminently achievable. If similar conditions also exist for other molecules, as may be expected, then this nondestructive technique would be readily applicable in many existing cold molecule experiments.

\section{Imaging Single Molecules}
\label{sec_summary_single}

A natural and impactful extension of this imaging scheme would be to enable the resolution of individually trapped molecules~\cite{NiSingleMolecule,doylecaf,KKNi-FBM}. For an individual point-like scatterer, such as a single molecule tightly confined to a lattice site or optical tweezer, the peak polarization rotation will be smaller than the values we have discussed for bulk molecular gases. This is because individual molecules will have a maximum effective optical density (OD), while the signal from a bulk gas can be boosted by the collective, integrated contribution of many molecules along the imaging direction. To compensate for this loss of collective OD enhancement, operation closer to resonance is required to attain degree-level rotations from single molecules. Furthermore, as discussed in Ref.~\cite{takahashisingleatom}, a high numerical aperture imaging system is required to enable the detection of individual particles.

We first consider the achievable polarization rotation signal under the most ideal conditions: utilizing a state-of-the-art imaging system with an NA of 0.8~\cite{BakrQGM} and operating on the more vertical \unit[884]{nm} transition. We additionally consider the case of ``parallel'' imaging, which reduces the amount of inelastic scattering by roughly a factor of two for the equivalent rotation signal. At a detuning of $\Delta=\unit[19]{\Gamma}$, a point-like scatterer would result in a peak polarization rotation of $\phi \approx 1.52^{\circ}$ under these conditions. While this degree of rotation is comparable to what has been used to detect single atoms~\cite{takahashisingleatom}, one also has to account for how much scattering can be tolerated for the molecules. For an imaging intensity of $I =$~\unit[0.02]{mW/cm\textsuperscript{2}}, as was considered in Table~\ref{tab:1}, this would result in an inelastic scattering rate of \unit[152]{Hz}.
%for ``parallel'' imaging on the \unit[884]{nm} transition.

We can restrict to an imaging time $\tau$ such that only one inelastic scattering event occurs and the molecule interacts with $N_p \approx I \tau \sigma/ hf$ probe photons, where $\sigma \approx \lambda^2/\pi$ is the off-resonant scattering cross-section for imaging light of wavelength $\lambda$ (frequency $f$). With this restriction, one finds that the maximum achievable signal-to-noise ratio (SNR) for shot-noise-limited performance, $\textrm{SNR}_{max} = \phi\sqrt{\eta N_p}$~\cite{HopeClose-Limit}, just barely exceeds 1 even if we assume a perfect efficiency $\eta$ for collection and detection. Under realistic conditions, the actual SNR will be reduced due to additional noise, reduced efficiency, by the use of imaging systems with more modest NA, and potentially by use of the ``perpendicular'' scheme or more lossy imaging transitions.
%As an example using the \unit[770]{nm} transition, an imaging system with numerical aperture of 0.65 and a detuning $\Delta=\unit[124]{\Gamma}$ yields a peak polarization rotation of \unit[2.59]{mrad}~$\approx 0.15^{\circ}$ (with the same $\gamma_\textrm{sc}$ as in Table~\ref{tab:1} for the same imaging intensity).
%While the measurement of mrad-level signals~\cite{budkersmallradmeasurement} is potentially feasible in cold molecule experiments, operation under such conditions may be aided by extensions based on cavity enhancement of the dispersive signal~\cite{Thompson-CavityAided} or modulation-based lock-in techniques.

To achieve the high SNRs necessary for high-fidelity detection, this dispersive imaging technique would thus have to be combined with, \textit{e.g.}, enhancement by a high-finesse optical cavity~\cite{Poldy,Thompson-CavityAided} or by the addition of repumping lasers, which would enable more scattering events prior to the loss of population to dark states~\cite{Kobayashi-Narrow}. In the latter case, repumping in a way that is commensurate with polarization-based dispersive imaging could be achieved by using $J=0$ ground state molecules. Dispersive imaging on narrow, nearly vertical transitions~\cite{Kobayashi-Narrow,Koto-Bloch-Narrow} could be enabled by the application of an electric field or optical fields, thereby breaking the degeneracy of the $J' = 1$ sublevels and inducing an anisotropic polarizability. 
%Alternatively, one may increase the signal by simply operating at smaller values of detuning, while having to tolerate an increased scattering rate $\gamma_\textrm{sc}$. For instance, a detuning of \unit[19]{$\Gamma$} on the \unit[770]{nm} transition is calculated to result in a peak polarization rotation of \unit[1]{\textsuperscript{$\circ$}} (for an NA of 0.65) with an inelastic scattering rate of \unit[629]{Hz}.
%At this detuning, the benefits of the \unit[884]{nm} transition's verticality become more pronounced, as it boasts the same polarization rotation with an inelastic scattering rate of \unit[286]{Hz}.
%Provided similar imaging conditions exist for other molecular species, including the bi-alkalis, it would seem that dispersive polarization-based imaging of individual molecules should already be within reach of current experiments.

\section{Discussion}
\label{sec_discussion}

In this paper we have presented a nondestructive technique for imaging excited rotational states of ultracold molecules using well known techniques from the toolbox of ultracold atoms. We described the anisotropic nature of excited rotational states and detailed how this can be translated to a measurable polarization rotation of a low intensity probe beam. For \narb and \rbcs we identified electronic transitions one might use to image the first rotational excited state and presented expected polarization rotations for conditions in the current state-of-the-art ultracold molecule experiments.

These capabilities will be especially important for systems that lack alternative detection schemes based on optical cycling transitions, such as hetero-nuclear bi-alkalis and homo-nuclear alkali dimers. The nondestructive nature of the proposed imaging method for bulk gases is well-suited to applications in the study of cold chemistry. For instance, the continuous monitoring of a single sample of molecules may allow for the study of losses by chemical reaction~\cite{Ospelkaus853}, while avoiding sensitivity to shot-to-shot variations in the number of molecules produced.

Through the incorporation of cavity-based enhancement of dispersive signals, the discussed approach has potential to impact fundamental physics, such as in the search for bosonic dark matter particles~\cite{Dark-Matter-Molecule}. One could continuously monitor molecular samples prepared in a ``dark'' rotational states that gives rise to no polarization rotation signal, looking for events in which population jumps to ``bright'' rotational states that yield a polarization rotation signal. Dispersive measurements aided by cavity enhancement could be utilized for measurement-based~\cite{Kuzmich-Squeeze,Kuzmich-QND,Appel-Squeeze,Thompson-CavityAided} and coherent~\cite{LerouxCavitySqueeze,Thompson-CavityAided} generation of squeezing of molecular rotation, which could then be transferred to alternate degrees of freedom to enable applications relevant to fundamental physics~\cite{acmedipole,Cairncross2017,Kobayashi2019}.

The extension of the proposed approach to the detection of individual molecules could be enabled either by cavity enhancement of the dispersive phase shift or by the addition of one or more repump lasers when utilizing narrow, ``vertical'' imaging transitions. These ideas are not fully developed as of yet and will require future studies. Such an extension would be of critical importance for QIS applications in fiducial state preparation~\cite{Endres1024,Barredo1021} and qubit readout.
%This technique is also naturally suited to studies of quantum magnetism based on rotational states of molecules as it can naturally be implemented on the pseudospin states (rotational ground and excited states). As an example, because this scheme is nondestructive, one could image population in an excited state, follow it with state inversion via adiabatic rapid passage, and repeat to image the second spin state.
Furthermore, this technique could enable effective quantum state preparation and high-fidelity detection in molecules, strengthening the relevance of molecules for use in quantum analog simulation~\cite{Lewenstein2007,Bloch2008} and precision measurement~\cite{Ludlow2015,Degen2017}.

\begin{acknowledgments}
The authors thank Ming Li, Wes Campbell, Kaden Hazzard, and Kang-Kuen Ni for insightful discussions and helpful feedback. All authors acknowledge support from the Air Force Office of Scientific Research Grant No. FA9550-19-1-0272. Q.G. and S.K. acknowledge funding from the Army Research Office Grant No. W911NF- 17-1-0563.  V.S. acknowledges support by the Air Force Office of Scientific Research Grant No. FA9550-18-1-0505 and Army Research Office Grant No W911NF-20-1-0013. M.H. and G.R.W. acknowledge support from the National Science Foundation Graduate Research Fellowship Program under Grant No. DGE1746047.
\end{acknowledgments}

\bibliographystyle{apsrev4-1}
\bibliography{molimg}

\newpage

\appendix
\section{Phase Shift for Parallel and Perpendicular Imaging} 
\label{append1}

In this section we derive Eq.~\ref{eq:phi}. The wave equation for an electric field $\vec{E}(\vec{r}, t)$ in an anisotropic medium is~\cite{Jackson1975}
\begin{equation}
\label{wave_equation}
\nabla(\nabla\cdot\vec{E}(\vec{r}, t)) - \nabla^2\vec{E}(\vec{r}, t) + \mu_0\epsilon_0\epsilon \frac{\partial^2}{\partial t^2} \vec{E}(\vec{r}, t) =0.
\end{equation}
Here, the relative permittivity $\epsilon$ should be understood as a tensor form. For diatomic molecules in a magnetic field along the $z$-direction, each eigenstate has a fixed projection of the total angular momentum along $z$. For molecules that occupy one of these eigenstates with no degeneracy, both the dynamic polarizability tensor $\alpha$ and the relative permittivity tensor $\epsilon$ are diagonal in the spherical basis $\hat{e}_{+1}=-(\hat{x}+i\hat{y})/\sqrt{2}$, $\hat{e}_{-1}=(\hat{x}-i\hat{y})/\sqrt{2}$, and $\hat{e}_0=\hat{z}$, where $\hat{x}$, $\hat{y}$, and $\hat{z}$ are the unit vectors of the three Cartesian coordinates. Given the condition $\rho c \alpha_{ii}\ll 1 (i = +, -, 0)$ where $\rho$ is the molecular number density, we can apply the Clausius–Mossotti relationship~\cite{Jackson1975} to relate the relative permittivity tensor to the dynamic polarizability tensor component by component via
\begin{eqnarray}
\label{relative_permittivity}
\epsilon_{ii}\approx 1+2\rho c\alpha_{ii}.
\end{eqnarray}
We look for a plane wave eigen-mode of $\vec{E}$,
\begin{eqnarray}
\label{plane_wave}
\vec{E}(\vec{r}, t) = \vec{F}\exp\left(\vec{k}\cdot\vec{r}-\omega t\right).
\end{eqnarray}
Plugging Eq.~\ref{plane_wave} into Eq.~\ref{wave_equation} and writing the equation in the spherical tensor basis, we have
\begin{widetext}
\begin{eqnarray}
\label{spherical_basis_wave_equation}
\begin{pmatrix}
\omega^2\epsilon_0\mu_0\epsilon_{++}-|\vec{k}|^2 + |k_{+1}|^2 & k_{+1}k_{-1}^* & k_{+1}k_0^*\\
k_{+1}^*k_{-1} & \omega^2\epsilon_0\mu_0\epsilon_{--}-|\vec{k}|^2 + |k_{-1}|^2 & k_{-1}k_0^*\\
k_{+1}^*k_0 & k_{-1}^*k_0 & \omega^2\epsilon_0\mu_0\epsilon_{00}-|\vec{k}|^2 + |k_{0}|^2
\end{pmatrix}
\begin{pmatrix}
F_{+1}\\
F_{-1}\\
F_{0}
\end{pmatrix}
=0.
\end{eqnarray}
\end{widetext}
Here, $k_{i}$ and $\vec{F}_{i}$ are the $i$-component of the vector $\vec{k}$ and $\vec{F}$, \textit{i.e.}, $k_{i} = \hat{e}^*_i\cdot\vec{k}$ and $F_i = \hat{e}_i^*\cdot\vec{F}$. For a fixed propagation direction $\vec{k}/|\vec{k}|$, the magnitude $|\vec{k}|$ of the wave vector of an eigen-mode is solved by setting the determinant of the $3\times 3$ matrix in Eq.~\ref{spherical_basis_wave_equation} equal to zero.

For the parallel imaging scheme, we have $\vec{k}=k_0\hat{e}_0$. In this case, the system is probed along one of the principle axes. We have the two eigen-mode solutions,
\begin{eqnarray}
\label{parallel_plus}
k_{0, +} = \frac{2\pi}{\lambda}\sqrt{\epsilon_{++}}  
\end{eqnarray}
and
\begin{eqnarray}
\label{parallel_minus}
k_{0, -} = \frac{2\pi}{\lambda}\sqrt{\epsilon_{--}},
\end{eqnarray}
where $\lambda$ is the probing laser wavelength. Using Eqs.~\ref{relative_permittivity},~\ref{parallel_plus}, and~\ref{parallel_minus}, the phase shift $\phi$ reads
\begin{align}
\label{paral_phase_shift}
\phi = &\left(k_{0, +}-k_{0, -}\right)L \\\nonumber
\approx &\frac{2\pi\rho c L}{\lambda}\left(\alpha_{++}-\alpha_{--}\right).
\end{align}

For the perpendicular imaging scheme, we have $\vec{k} = -k_{y}/(\sqrt{2}i)(\hat{e}_{+1}+\hat{e}_{-1})$. The two eigen-mode solutions read
\begin{eqnarray}
\label{perp_plus}
k_{y,x}=\frac{2\pi}{\lambda}\sqrt{\frac{2\epsilon_{++}\epsilon_{--}}{\epsilon_{++}+\epsilon_{--}}}
\end{eqnarray}
and
\begin{eqnarray}
\label{perp_minus}
k_{y,z}=\frac{2\pi}{\lambda}\sqrt{\epsilon_{00}}.
\end{eqnarray}
The solutions in Eqs.~\ref{perp_plus} and~\ref{perp_minus} correspond to a plane wave with the dominant polarization along the $x$- and $z$-directions, respectively. The phase shift $\phi$ for the perpendicular imaging scheme reads
\begin{align}
\label{perp_phase_shift}
\phi = &\left(k_{y, z}-k_{y, x}\right)L \\\nonumber
\approx & \frac{2\pi\rho c L}{\lambda}\left(\alpha_{00}-\frac{\alpha_{--}+\alpha_{++}}{2}\right).
\end{align}
In obtaining Eq.~\ref{perp_phase_shift}, we Taylor expand $k_{y,x}$ in Eq.~\ref{perp_plus} and neglect all the higher order terms of $\left(\rho c\alpha_{ii}\right)^n$ with $n>1$.

For the diatomic molecule in a magnetic field along the $z$-direction, the dynamic polarizability tensor of an eigenstate in the Cartesian coordinate have $\alpha_{xx}=\alpha_{yy}$ and $\alpha_{xz}=\alpha_{zx}=\alpha_{yz}=\alpha_{zy}=0$. Since the trace of a tensor is independent of the representation and the coupling between the $x,y$ degrees of freedom and the $z$ degree of freedom vanishes, 
we have
\begin{eqnarray}
\label{xy}
\alpha_{++}+\alpha_{--} = 2\alpha_{xx}
\end{eqnarray}
and
\begin{eqnarray}
\label{z}
\alpha_{00} = \alpha_{zz}.
\end{eqnarray}
Based on Eqs.~\ref{xy} and~\ref{z}, the phase shift $\phi$ in Eq.~\ref{perp_phase_shift} is
\begin{eqnarray}
\label{perp_phase_shift_2}
\phi \approx 
\frac{2\pi\rho c L}{\lambda}\left(\alpha_{zz}-\alpha_{xx}\right)
\end{eqnarray}
\section{Calculation of the Eigenstates of the \texorpdfstring{A$^1\Sigma^+$-b$^3\Pi_0$}{A1E+-b3P0} System}
\label{appendix_new}

To calculate the dynamic polarizabilities $\alpha_{zz}(\omega)$ and $\alpha_{xx}(\omega)$, we sum up contributions from the ro-vibrational and scattering states of ground and excited electronic states using the approach developed in Ref.~\cite{Kotochigova2006}. For the strongly coupled A$^1\Sigma^{+}$-b$^3\Pi_0$ system we rely on the electronic potentials surfaces, transition dipole moments, and spin-orbit coupling functions of Ref.~\cite{Decenko2007}. The relevant $R$-dependent electric transition dipole moments $d_{f\leftarrow i}(R)$ between the pairs $(i,f)=(\text{X}^1\Sigma^+,\,\text{A}^1\Sigma^+)\ \text{and}\ (i,f)=(\text{a}^3\Sigma^+,\,\text{b}^3\Pi$) have been taken from Refs.~\cite{Decenko2007,Aymar2007}. Transitions between the pairs X$^1\Sigma^{+}$-b$^3\Pi$  and a$^3\Sigma^{+}$-A$^1\Sigma^+$ of non-relativistic states are dipole forbidden. Moreover, the electric dipole moment operator only couples basis states with the same nuclear spin projection quantum numbers. For distant non-resonant electronic states, not shown in Fig.~\ref{fig:pots}, we use the potentials and transition dipole moments of Ref.~\cite{Romain2017}.

In this work, we are interested in the dynamic polarizabilities and the photon scattering rate near the resonance transitions to $J'=0$ target states of the A$^1\Sigma^+$-b$^3\Pi_0$ system. The rotational states with $J'=0$ only exist for electronic states with projection quantum number $\Omega^\sigma=0^\pm$, where $\Omega$ is the projection of the total electron spin and angular momentum on the internuclear axis and $\sigma=\pm$ denotes a reflection symmetry. In alkali-metal dimers only $\Omega'=0^+$ states can be excited from the X$^1\Sigma^{+}$ ground state. To further specify the target state we assembled the relevant $0^+$ potentials of \narb from Refs.~\cite{Pashov2005,Decenko2007}. Fig.~\ref{fig:pots} shows the X$^1\Sigma^{+}$ potential and the energetically lowest two $\Omega'=0^+$ relativistic potentials dissociating to atom pair states with one atom electronically excited. The latter two potentials have been obtained by diagonalizing at each $R$ a $2\times2$ potential matrix containing the non-relativistic A$^1\Sigma^{+}$ and b$^3\Pi$ electronic potentials coupled and shifted by an $R$-dependent relativistic spin-orbit interaction. For completeness, Fig.~\ref{fig:pots} also shows the a$^3\Sigma^+$ potential from Ref.~\cite{Pashov2005} as the b$^3\Pi_0$ state can decay into this state by spontaneous emission. This process contributes to $\gamma_n$, the natural linewidth.

The couplings in the $J'=0$ A$^1\Sigma^{+}$-b$^3\Pi_0$ system are sufficiently strong such that a quantitative representation of the molecular vibration requires a coupled-channel calculation starting from the non-relativistic basis of $|{\rm A}^1\Sigma^{+}\rangle$ and $|{\rm b}^3\Pi_0\rangle$ states, their potentials, and spin-orbit induced coupling. The normalized $J'=0$ target vibrational wavefunctions are given by
\begin{eqnarray}
  |\psi_{{\rm t},n'}\rangle&=&\frac{1}{\sqrt{4\pi}}
  \left( f_{{\rm A},n'}(R) |\text{A}^1\Sigma^{+}\rangle +
f_{{\rm b},n'}(R) |\text{b}^3\Pi_0\rangle \right) 
 \nonumber\\
  && \quad\quad\quad  \times \, | i_{\rm Na} m'_{\rm Na}\rangle | i_{\rm Rb} m'_{\rm Rb}\rangle \,,
  \label{eq:psi_exc}
\end{eqnarray}
where the functions $f_{{\rm A},n'}(R)$ and $f_{{\rm b},n'}(R)$ are obtained from the coupled-channel calculation and index $n'=0,\,1,\,2,\dots$ labels eigenstates by order of their eigenenergies. For $J'=0$ states the nuclear spin wavefunction is separable from that of the electrons and molecular rotation. The energy of two energetically nearest neighbor states with different $m'_{\rm Na}$ and $m'_{\rm Rb}$ are spaced by the nuclear Zeeman interaction and of order \unit[$h\times0.1$]{MHz} for our magnetic field strength. The quantities $a_{s,n'}=\int_0^\infty r^2 {\rm d}r |f_{s,n'}(R)|^2$ are the admixtures of eigenstate $n'$ in electronic components $s=\text{A or B}$. For ease of notation we suppress the rotational and nuclear spin quantum numbers in denoting target states $|\psi_{{\rm t},n'}\rangle$. Effects of Coriolis-induced coupling to ro-vibrational levels of $\Omega'=0$\textsuperscript{--}, 1, and 2 potentials of the b$^3\Pi$ state are negligible for our purposes.

\section{Derivation of Differential Transition Width} 
\label{append_derive_gamma}

In this section we argue that Eqs.~\ref{eq_diff_linewidth}-\ref{eq:transition_width_2} offer a good approximation to the differential transition width $\Gamma$. First we note that the superposition of nuclear spin states in $|\varphi_{\rm perp}\rangle $ in Eq.~\ref{eq:imag_state_perp} leads to contributions to $\Delta\alpha(\omega)$ from two nearly-degenerate target states with the same state label $n'$ and quantum number $J'=0$ and $M'=0$, but different  nuclear spin projections $m'_{\rm Rb}$ of  $^{87}$Rb. At \unit[$B=335.6$]{G} these two target states are split by $h\times$\unit[0.1]{MHz}.  We find that the value is on the order of or smaller than the natural linewidth of eigenstates of the A$^1\Sigma^{+}$-b$^3\Pi_0$ complex. In fact, as the superposition of states in $|\varphi_{\rm perp}\rangle$ also corresponds to a superposition of states with different rotational projection quantum numbers $M$, the $M=0$ component contributes to $\alpha_{zz}(\omega)$ and the $M=1$ component to  $\alpha_{xx}(\omega)$. Then for detunings $|\Delta|\gg\gamma_n$, we can neglect the \unit[$h\times0.1$]{MHz} energy difference and define the differential transition width as in Eq.~\ref{eq_diff_linewidth}. Then for the $n'$-th $J'=0$ ro-vibrational target state $|\psi_{{\rm t},n'}\rangle$ of the A$^1\Sigma^+$-b$^3\Pi_0$ system, we arrive at Eq.~\ref{eq:transition_width_2}, where the vibrational matrix element is:
\begin{eqnarray}
\label{eq:mu}
\mu_{n'} = \int_0^\infty R^2 {\rm d}R \, f_{{\rm A},n'}(R) d_{A\leftarrow X}(R) \varphi_{\rm perp}(R)
\end{eqnarray}
and $f_{{\rm A},n'}(R)$ and $\varphi_{\rm perp}(R)$ the radial wavefunction of the A$^1\Sigma^+$ component of $|\psi_{{\rm t},n'}\rangle$ and the radial wavefunction of the imaging state $|\varphi_{\rm perp}\rangle$, respectively [see Appendix~\ref{appendix_new}].

\section{Target States for Parallel Imaging Scheme}
\label{appendparallel}
\begin{figure}
    \includegraphics[trim = 0.8cm 10cm 0cm 3cm, clip,width=0.95\textwidth]{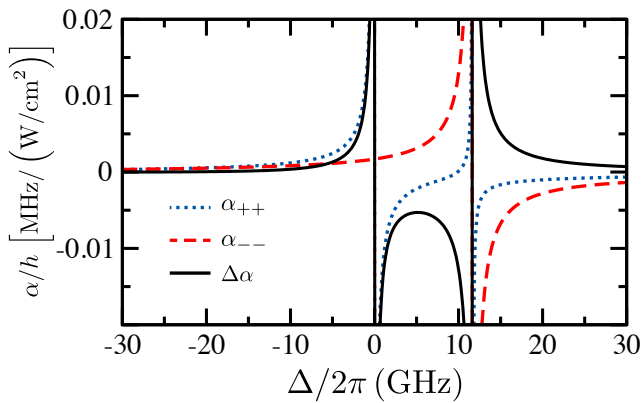}
     \vspace*{-0.8cm}
    \caption{
        Dynamic polarizabilities $\alpha_{++}(\omega)$ and $\alpha_{--}(\omega)$ and corresponding differential dynamic polarizability $\Delta\alpha(\omega)$ for parallel imaging based on the $|\varphi_{\rm paral}\rangle$ imaging state near the 770 nm transition of the $^{23}$Na$^{87}$Rb molecule shown in Fig.~\ref{fig:pots}. We use \unit[$B=335.6$]{G}.}
    \label{fig:circular}
\end{figure}

We can also probe the molecular system with light propagating parallel to the magnetic field direction, the so-called ``parallel'' probing scheme.  In such a case, the probe laser is linearly polarized with the polarization lying in the plane perpendicular to the $B$-field. We selected one of the higher energy hyperfine-Zeeman states as our imaging state for parallel imaging. This level is the only state with $M_{\rm tot}=-4$ and is thus given by
\begin{equation}
\left|\varphi_{\text{paral}}\right\rangle=\left|\text{X}^1\Sigma^+;\nolinefrac{v=0, J=1,M=-1}{m_{\text{Na}}=-3/2, m_{\text{Rb}}=-3/2}\right\rangle\,.
\label{eq:imag_state_par}
\end{equation}
It has the second highest energy of the ${v=0,J=1}$ X$^1\Sigma^+$ hyperfine states. The only $M_{\rm tot}=+4$ state can be used as an imaging state as well. For these ``circularly polarized'' ${|M|=1}$ states, the relevant differential polarizability is
\begin{equation}
\Delta\alpha(\omega)=\alpha_{++}(\omega)-\alpha_{--}(\omega),
\end{equation}
where $\alpha_{++}$ and $\alpha_{--}$ are spherical tensor components of the rank-2 dynamic polarizability tensor. This differential polarizability relates to a circular birefringence of the molecules, which will give rise to direct rotation of the probe beam's linear polarization vector.

Figure~\ref{fig:circular} shows the dynamic polarizabilities $\alpha_{++}(\omega)$, $\alpha_{--}(\omega)$, and $\Delta\alpha(\omega)$ for the \unit[770]{nm} transition to the ${n'=39}$ state of the A$^1\Sigma^{+}$-b$^3\Pi_0$ complex. The poles at \unit[$\Delta=0$]{GHz} and \unit[11.65]{GHz} correspond to  resonant transitions to $J'=0$ and $J'=2$ rotational states, respectively. The $J'=0$ pole is absent in the curve for $\alpha_{--}(\omega)$ as only $M'=-2$ states are accessible
for this polarization tensor component.

In the parallel probing scheme, the differential transition width $\Gamma$ for the $J'=0$ transition is larger than that for the perpendicular probing scheme. In fact, the parallel differential transition width is $(c_1^2 - c_2^2/2)^{-1}=1.44$ times larger for all eigenstates $n'$, leading to differential transition widths of $2\pi\times$\unit[5.7]{kHz}, $2\pi\times$\unit[274]{kHz}, and $2\pi\times$\unit[27.8]{kHz} for the \unit[884]{nm}, \unit[791]{nm}, \unit[770]{nm} transitions, respectively. The natural linewidths for the parallel probing scheme are the same as those for the perpendicular probing scheme. Thus, for the same detuning, the parallel probing scheme gives a slightly larger phase difference $\phi$ than the perpendicular probing scheme.

Finally, we note that the $M_{\rm tot}= \pm 4$ states, which are most ideal for the parallel probing scheme, can also be utilized for the perpendicular probing scheme. In this case, the pole in the $\alpha_{zz}$ polarizability vanishes near the $J = 1$ to $J' = 0$ transition, while $\alpha_{xx}$ features a prominent pole, as the linear polarization along the $\hat{x}$ axis can drive both $\sigma_+$ and $\sigma_-$ transitions. While the transition widths for these states in the perpendicular scheme are reduced by a factor of 2 from the values they take in the parallel scheme, they will nevertheless give rise to appreciable polarization rotation. More generally, a birefringent response should be possible for \textit{any} state with $J \neq 0$ in either imaging scheme, while for each approach particular states will provide the largest possible rotation signals.

\section{Imaging \rbcs Molecules}
\label{append3}

\begin{figure}
\includegraphics[scale=0.6]{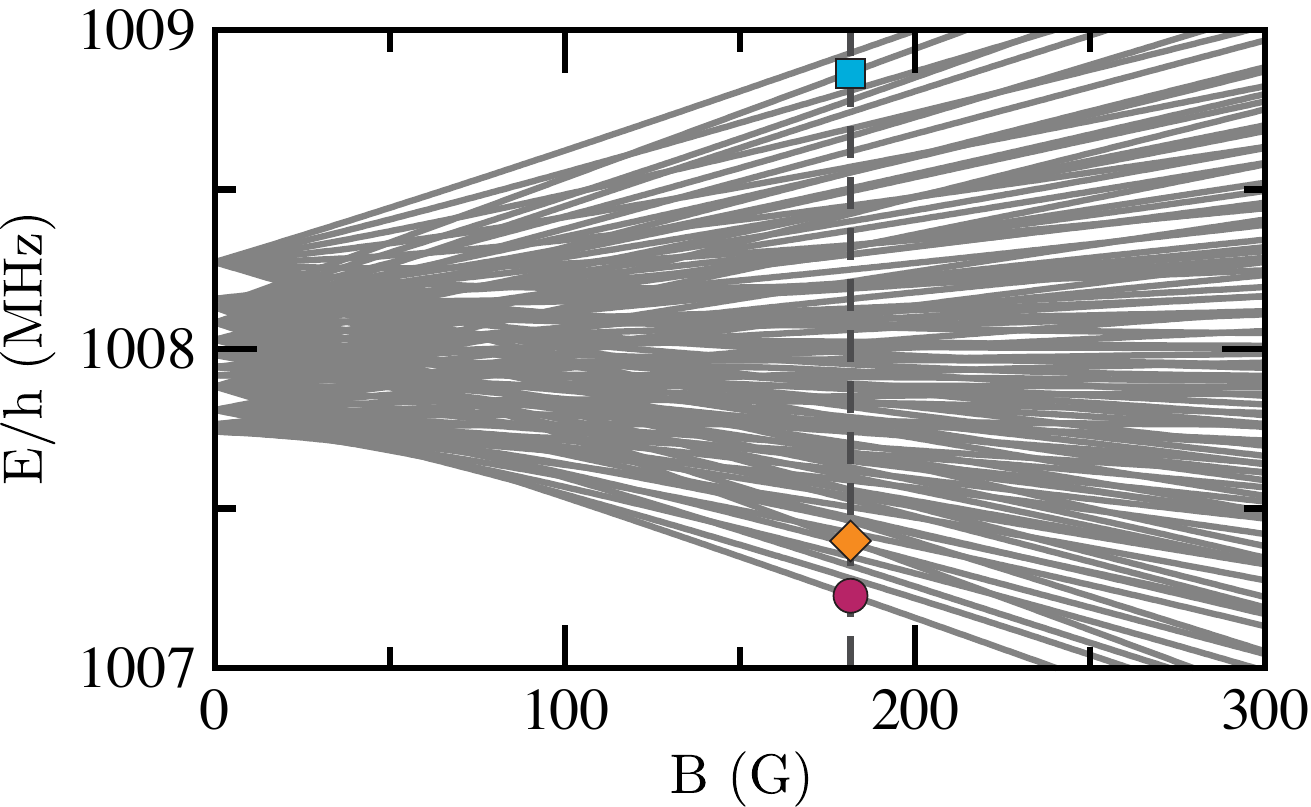}
\caption{The 96 hyperfine and Zeeman energy levels for the $v=0,J=1$ manifold of  the X$^1\Sigma^+$ state of $^{87}$Rb$^{133}$Cs. The vertical line indicates  magnetic field $B=182$ G.  At this magnetic field strength the level with a magenta dot corresponds to the optimal  state for the perpendicular probing
scheme, while the cyan square and orange diamond correspond to the best states for the parallel probing scheme.
}
\label{fig:zeeman_RbCs}
\end{figure}

In this section, we analyze nondestructive imaging of the ${v=0,J=1}$ ro-vibrational level of the X$^1\Sigma^+$ state of \rbcs. Ultracold \rbcs, another bi-alkali molecule, has been created using  STIRAP from cold atom gases close to an interspecies Feshbach resonance near $B=\unit[182]{\text{G}}$ \cite{Molony2014}. Fig.~\ref{fig:zeeman_RbCs} shows the 96 hyperfine/Zeeman eigenenergies of the $v=0, J=1$ level as a function of magnetic field strength $B$. The nuclear spins of \Rb and \Cs are $3/2$ and $7/2$, respectively, and we use the nuclear quadrupole moments and nuclear $g$ factors from Ref.~\cite{Aldegunde2008}.

We determine dynamic polarizabilities at $B=182$\,G, indicated in Fig.~\ref{fig:zeeman_RbCs}, close to the Feshbach resonance location used by Ref.~\cite{Molony2014}. For the perpendicular imaging scheme, we
use the imaging state
\begin{eqnarray}
\label{eq:imaging_state_RbCs}
|\varphi_{\text{perp}}\rangle &=& c_0 \left| {\rm X}^1\Sigma^+;\nolinefrac{v=0, J=1, M=0}{m_{\text{Rb}}=3/2, m_{\text{Cs}}=7/2}\right\rangle \\
&& \quad\quad+\,c_1 \left| {\rm X}^1\Sigma^+;\nolinefrac{v=0, J=1, M=1}{m_{\text{Rb}}=1/2, m_{\text{Cs}}=7/2}\right\rangle
\nonumber
\end{eqnarray}
with $c_0=0.925$ and $c_1 = 0.374$. It is the energetically lowest $J=1$ hyperfine state and, again, has the largest $M=0$ contribution of all $J=1$ hyperfine states. For parallel imaging, we consider the $M=-1$ state 
\begin{equation}
\label{eq:imaging_state_RbCs_para}
|\varphi_{\text{paral}}\rangle = \left|{\rm X}^1\Sigma^+; \nolinefrac{v=0, J=1, M=-1}{m_{\text{Rb}}=-3/2, m_{\text{Cs}}=-7/2}\right\rangle
\end{equation}
with stretched nuclear Zeeman states such that $|m_{\rm Rb}|=i_{\rm Rb}$ and $|m_{\rm Cs}|=i_{\rm Cs}$. The polarizability for the hyperfine state with all projection quantum numbers of opposite sign is the same as that for $|\varphi_{\text{paral}}\rangle$. All three states are marked in Fig.~\ref{fig:zeeman_RbCs}.

\begin{figure}
\includegraphics[trim = 20 45 0 80, clip, width=1\textwidth]{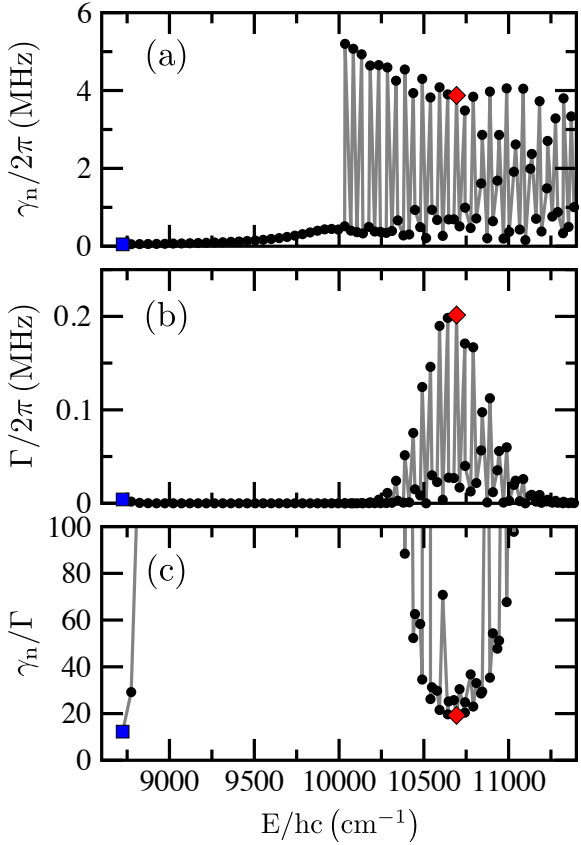}
\vspace*{-0.5cm}
\caption{
The natural linewidth $\gamma_{\rm n}$ \textbf{(a)} of eigenstates of the $J'=0$ A$^1\Sigma^{+}$-b$^3\Pi_0$ complex of $^{87}$Rb$^{133}$Cs and their differential transition width $\Gamma$ \textbf{(b)} from the ground $v=0,J=1$ X$^1\Sigma^+$ imaging state $|\varphi_{\rm perp}\rangle$ for perpendicular imaging as functions of transition energy $E$. The transition energies are relative to the energy of $|\varphi_{\rm perp}\rangle$ and the applied magnetic field is $B=182$~G. Panel \textbf{(c)} shows the ratio $\gamma_{\rm n}/\Gamma$ as a function of transition energy. The blue square and red diamond correspond to the transitions with the smallest $\gamma_{\rm n}/\Gamma$ and the largest $\Gamma$, respectively. The values of $\gamma_{\rm n}/\Gamma$ for the blue square and the red diamond are 12 and 19, respectively. The values of $\Gamma$ for the blue square and the red diamond are \unit[$2\pi\times 4.05$]{kHz} and \unit[$2\pi\times 201$]{kHz}, respectively.
}
\label{fig:RbCs}
\end{figure}

For the imaging state, we use the X$^1\Sigma^+$ potential from Refs.~\cite{Docenko2011,Mario2016}. For the target A$^1\Sigma^+$-b$^3\Pi_0$ complex, we use the potentials and spin-orbit matrix element from Ref.~\cite{Docenko2010}. To calculate the natural linewidths of the A$^1\Sigma^+$-b$^3\Pi_0$ complex, the spontaneous decay to the a$^3\Sigma^+$ potential is included. The a$^3\Sigma^+$ potential is taken from Refs.~\cite{Docenko2011,Mario2016}. Other excited electronic potentials have been taken from Ref.~\cite{Dulieu2017}. Finally, transition electric dipole moments are taken from Refs.~\cite{Zuters2013,Romain2017}.

Figure~\ref{fig:RbCs} shows the natural linewidth $\gamma_{\rm n}$, the differential transition width $\Gamma=\Gamma_{zz}-\Gamma_{xx}$, and the ratio $\gamma_{\rm n}/\Gamma$ as functions of the transition energy from $|\varphi_{\rm perp}\rangle$ to ${J'=0, M'=0}$ target eigenstates of the coupled A$^1\Sigma^+$-b$^3\Pi_0$ complex for the perpendicular imaging scheme. The natural linewidths in Fig.~\pref[a]{fig:RbCs} are much smaller than $2\pi\times \unit[1]{\text{MHz}}$ for target states with transition energies $E$ less than $hc\times\unit[10\,035]{\text{cm}^{-1}}$. These eigenstates have energies below the minimum of the A$^1\Sigma^+$ potentials and, thus, have a large b$^3\Pi_0$ admixture and small natural linewidths. For $E/hc >\unit[10\,035]{\text{cm}^{-1}}$, the ordering of the eigenenergies alternate between the one with dominant A$^1\Sigma^+$ and the one with dominant b$^3\Pi_0$ admixture leading to alternating large and small natural linewidths.

The differential transition width $\Gamma$ for the perpendicular imaging scheme is shown in Fig.~\pref[b]{fig:RbCs}. The values of $\Gamma$ are less than \unit[$2\pi\times 5$]{kHz} for $\unit[E/hc<10\,035]{\text{cm}^{-1}}$ due to the forbidden nature of the dipole transitions from the X$^1\Sigma^+$ state to the b$^3\Pi_0$ state. For target states with $\unit[10\,035]{\text{cm}^{-1}}~<~E/hc~<~\unit[11\,400]{\text{cm}^{-1}}$, the differential transition widths are positive and oscillatory with a Gaussian envelope. The largest $\Gamma$ is \unit[$2\pi\times$0.201]{MHz} for the target state with a transition wave length of \unit[935]{nm}. For the parallel imaging scheme, not shown, the differential transition widths are 1.27 times larger than those in Fig.~\pref[b]{fig:RbCs}, as again follows from the coefficients $c_i$ in Eq.~\ref{eq:imaging_state_RbCs}. 

From Fig.~\pref[c]{fig:RbCs} we see that the ratio between the natural linewidth and the differential transition width is larger than 100 for the target states with $E/hc<\unit[10\,035]{\text{cm}^{-1}}$ with the exception of the first two. For $\unit[10\,035]{\text{cm}^{-1}}~<~E/hc~<~\unit[11\,400]{\text{cm}^{-1}}$, some of the ratios are smaller than 100. The smallest ratio is 12 and occurs for the transition to the bottom of the b$^3\Pi_0$ potential. The ratio for the second lowest eigenstate is 29. The few transitions around and including the one with the largest transition width have the ratios close to 19. Consequently, the two energetically lowest eigenstate and quite a few eigenstates with transition wavelengths near \unit[935]{nm} can be used for nondestructive imaging. 

Since the smallest ratio $\gamma_{\rm n}/\Gamma$ for the \narb systems is a little bit smaller than the \rbcs systems, imaging of \narb molecules will be less destructive than that of \rbcs systems if the most vertical transition to the energetically lowest eigenstate of the A$^1\Sigma^+$-b$^3\Pi_0$ complex is used.

We note that the STIRAP ``dump'' transition for \rbcs relates to a transition to the b$^3\Pi_1$ excited state, outside of the range of transitions we have explored. For \narb it was determined that the STIRAP ``dump'' transition can be readily applied to nondestructive dispersive imaging of bulk molecular gases. It remains to be determined if such a convenient choices for an imaging laser could be applicable for the other bi-alkali species, and more generally for other molecules produced by STIRAP. 

\end{document}